\documentclass[5p,twocolumn]{elsarticle}

\usepackage{amssymb,amsthm}
\usepackage{multirow}
\usepackage{flushend}
\usepackage[utf8x]{inputenc}

\begin{document}

\begin{frontmatter}



\title{Low-temperature thermal expansion of the topological material candidates $\beta$-PtBi$_2$ and $\beta$-Bi$_2$Pd}


\author[CAB]{V. F. Correa\corref{mycorrespondingauthor}}
\cortext[mycorrespondingauthor]{Corresponding author}
\ead{victor.correa@cab.cnea.gov.ar}

\author[CAB,IFW]{P. Pedrazzini}

\author[CAB]{D. G. Franco}

\author[UJI]{A. J. Rosa}

\author[IFW,TU]{B. Rubrecht}

\author[CAB]{N. Haberkorn}

\address[CAB]{Centro At\'omico Bariloche and Instituto Balseiro, CNEA, CONICET and U. N. de Cuyo, 8400 San Carlos de Bariloche, Argentina}

\address[IFW]{Leibniz IFW Dresden, Helmholtzstra\ss e 20, Dresden, Germany}

\address[UJI]{Institute of Advanced Materials (INAM), Universitat Jaume I, 12006 Castell\'o, Spain}

\address[TU]{Institute for Solid State and Material Physics, TU Dresden, 01062 Dresden, Germany}

%
						%

											%
%

\begin{abstract}

We report on the low-temperature ($T <$ 120 K) thermal expansion of the bismuth-based topological semimetal $\beta$-PtBi$_2$ and topological superconductor $\beta$-Bi$_2$Pd candidates. 
The linear thermal-expansion coefficient of tetragonal $\beta$-Bi$_2$Pd shows a pronounced anisotropy between the $a$- and $c$-axis while the volume thermal-expansion coefficient $\alpha_V$($\beta$-Bi$_2$Pd) is considerable larger than $\alpha_V$($\beta$-PtBi$_2$). 
The coefficient $\alpha_V$($\beta$-PtBi$_2$) nearly matches the experimental specific heat, from which a Debye temperature $\theta_D =$ 199 K is obtained. On the other hand, $\alpha_V$($\beta$-Bi$_2$Pd) reasonably fits the Debye model with $\theta_D =$ 138 K, extracted from the low-temperature specific heat.
An almost constant Gr\"uneisen parameter $\Gamma \approx$ 2 is obtained for both compounds.
No magnetostriction is observed in any of both compounds up to $\mu_0 H =$ 16 T.
We compare our results with other Bi-based topological materials.

\end{abstract}

%


\end{frontmatter}


\section{Introduction}
\label{intro}

Topological matter (TM) has received a considerable interest in recent years \cite{Hasan2010,Qi2011,Sato2017,Yan2017,Armitage2018}. Unlike traditional states of matter, the character of topological phases are not determined by any physicochemical property of the underlying material but, ultimately, by the presence or lack of some internal symmetry.
The very peculiarity of the electronic-topological materials is that their bulk properties differ from those at the surface \cite{Hasan2010}.  

In topological insulators (TI), a crossing between the valence and conduction bands occurs. The residual spin-orbit interaction may open a considerable full gap at the Fermi level, ultimately giving rise to metallic surface states while the bulk material remains insulating \cite{Hasan2010}. The metallic character of the surface cannot be destroyed by passivation or partial doping. It depends only on certain symmetries, particularly the time-inversion symmetry \cite{Hasan2010}: they are said to be symmetry-protected surface states.
On the other hand, in topological semimetals (TSM) the gap at the Fermi level is not complete. At high-symmetry locations of the reciprocal space (Dirac points) gapless touching bands can result in linearly dispersive energies, as it occurs in graphene. Unlike graphene, however, TSM are three-dimensional systems. Again, distinctive metallic topology-protected surface states appear \cite{Yan2017}.

Other gapped systems can also show robust metallic surface states. This is the case of topological superconductors (TSC), where the bulk of the system remains gapped and superconductor while the surface shows symmetry-protected metallic states filled up with very particular quasi-particles known as Majorana fermions \cite{Qi2011,Sato2017,Leijnse2012}.

The list of proposed topological compounds is quite extensive. The first experimental realization of a TI was in HgTe/CdTe quantum wells \cite{Bernevig2006,Konig2007} while Bi$_{1-x}$Sb$_x$ \cite{Hsieh2008}, Bi$_2$Se$_3$ \cite{Zhang2009,Xia2009}, Bi$_2$Te$_3$ \cite{Zhang2009}, Sb$_2$Te$_3$ \cite{Zhang2009}, and the rare earth Heusler compounds LnPtBi (Ln=Y, La, and Lu) \cite{Chadov2010,Lin2010,Xiao2010} can be listed as intrinsic TI.
On the other hand, some TSM candidates are WTe$_2$ \cite{Ali2014}, MoTe$_2$ \cite{Sun2015}, PtBi$_2$ \cite{Gao2017}, TaAs \cite{Lv2015}, TaP \cite{Xu2015}, NbAs \cite{Fang2019}, NbP \cite{Shekhar2015}, and Cd$_3$As$_2$ \cite{Liu2014} among many others.
Although far from conclusive, the first proposed TSC was Sr$_2$RuO$_4$ \cite{Mackenzie2013}. Some other compounds with experimental results supporting topological superconductivity are Cu$_x$Bi$_2$Se$_3$ \cite{Hor2010}, Sn$_{1-x}$In$_x$Te \cite{Sasaki2012}, BiPd \cite{Benia2016}, and TaC \cite{Shang2020}.
An almost universal feature of topological compounds is the presence of a metalloid, a chalcogen or close element (P, Sn, Bi) in their unit formula.

Besides the natural interest in TM from the purely theoretical and fundamental side, some topological systems are also promising candidates for applications in spintronics and high-speed electronics \cite{Wang2016}. Most experimental and theoretical studies have been directed to the low-temperature electronic properties. On the other hand, lattice studies have received poor or null attention. A very small electron-lattice coupling explains this fact. Nonetheless, and mostly with prospective applications in mind, it is important to know the lattice behavior as the temperature changes. Large differential thermal expansions should always be avoided in devices.

In this work we report on the low-temperature ($T <$ 120 K) linear and volume thermal expansion measurements of the bismuth-based binary compounds $\beta$-PtBi$_2$ and $\beta$-Bi$_2$Pd. The former is a TSM candidate that shows one of the largest non-saturating transverse magnetoresistances \cite{Gao2017}. The latter is a superconductor with one of the highest critical temperatures ($T_c$ = 5.4 K) among the TSC candidates \cite{Imai2012}. 
We show that although both systems have comparable unit-cell volumes, their volume thermal-expansion coefficients differ substantially but basically track the specific heat. The linear thermal-expansion coefficient of tetragonal $\beta$-Bi$_2$ is markedly anisotropic.

The rest of this manuscript is organized as follows. Section \ref{II} is devoted to the crystal synthesis and characterization of our samples. Section \ref{III} presents the thermal-expansion results and a comparison with other bismuth-based systems in the literature. Finally, in section \ref{IV}, we present the conclusions.

\section{Crystal synthesis and characterization}
\label{II}

\begin{figure*}[t]
\centering
\includegraphics[width=\textwidth]{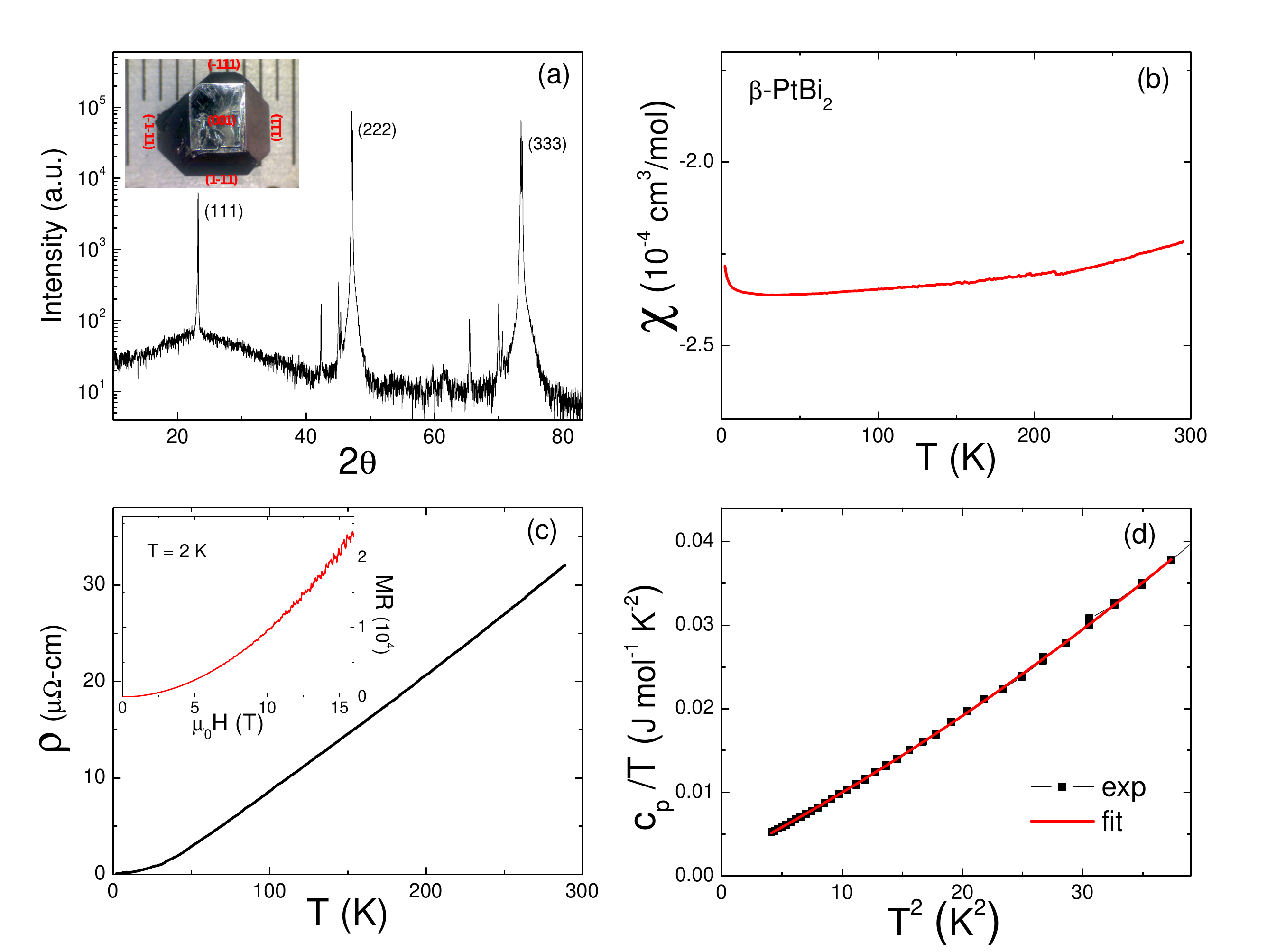}
\caption[]{(a) X-ray diffraction scan of a $\beta$-PtBi$_2$ crystal corresponding to the (hhh) family planes. Inset: crystal image with facets labeled. (b) Temperature dependence of the magnetic susceptibility under an applied field $H$ = 5000 Oe. (c) Temperature dependence of the electrical resistivity. Inset: magnetoresistance $\left( R(H)-R(0) \right) / R(0)$ as a function of the applied magnetic field $H$. (d) Specific heat divided by temperature versus square temperature along with a polynomial fit (see text for details).}
\label{fig1}
\end{figure*}

Single crystals of $\beta$-PtBi$_2$ and $\beta$-Bi$_2$Pd were grown by the self-flux technique. Initial mixtures of pure elements (Bi: $>$ 99.999$\%$; Pt: $>$ 99.98$\%$; Pd: $>$ 99.9$\%$) are placed in previously evacuated quartz tubes that are then sealed after incorporating a small amount of Ar. Mixtures are homogeneized at high temperature and then cooled down following specific ramps. Both $\beta$ phases are metastable at room temperature so the initial mixture must be quenched to obtain good crystalline samples. Crystals decompose when temperature is raised. In particular, $\beta$-Bi$_2$Pd starts to transform into the stable $\alpha$-Bi$_2$Pd phase at temperatures as low as 60 $^{\circ}$C, which critically affects the sample quality \cite{Haberkorn2022}.
Energy-dispersive x-ray spectroscopy (EDS) and Cu $K_{\alpha}$ x-ray diffraction (XRD) scans confirm the correct stoichiometry and crystal structure of the samples. $\beta$-PtBi$_2$ crystallizes in the cubic $Pa\bar{3}$ structure, while $\beta$-Bi$_2$Pd does it in the tetragonal $I4/mmm$ structure.

Magnetization (M) experiments were performed in a Quantum Design MPMS magnetometer while a standard four-probe setup was used in the electrical transport measurements. Specific heat was measured in a PPMS using a standard relaxation technique.

\begin{figure*}[t]
\centering
\includegraphics[width=\textwidth]{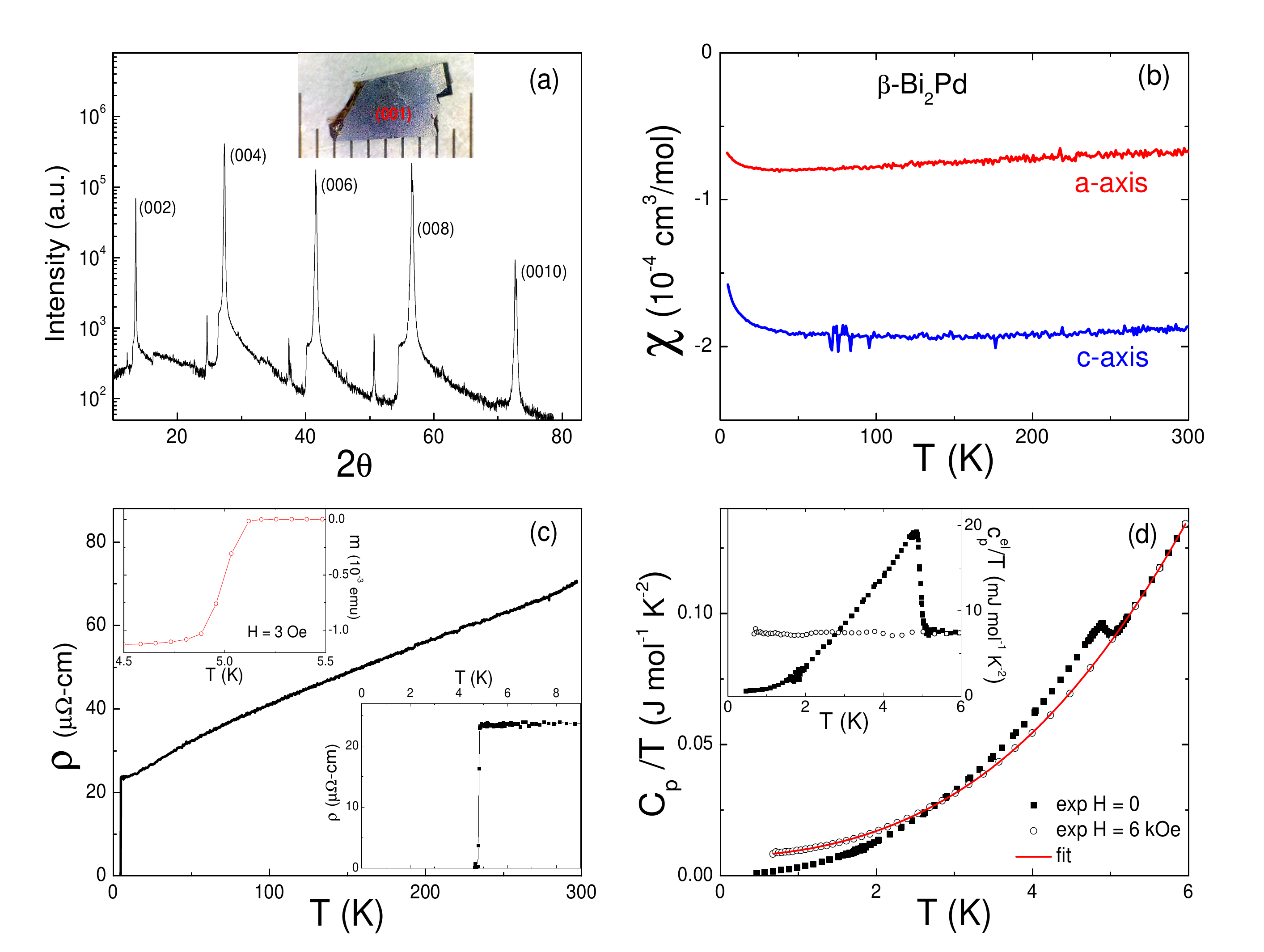}
\caption[]{(a) X-ray diffraction scan of a $\beta$-Bi$_2$Pd crystal corresponding to the (00l) family planes. Inset: crystal image. (b) Temperature dependence of the magnetic susceptibility under an applied field $H$ = 10000 Oe. (c) Temperature dependence of the electrical resistivity. Lower inset: superconducting transition from resistivity at $T_c =$ 5.1 K. Upper inset: Magnetic moment as a function of temperature showing a Meissner state below $T_c =$ 5.1 K. (d) Specific heat divided by temperature versus temperature at zero field and $H =$ 6 kOe along with a polynomial fit (see text for details). Inset: Electronic contribution to the specific heat $c_p^{el}$.}
\label{fig2}
\end{figure*}

\subsection{$\beta$-PtBi$_2$}
\label{Pt}

The initial mixture (Pt:Bi = 3-4:97-96) is homogeneized at 500 $^{\circ}$C for 4-10 hours, then slowly cooled down to 300 $^{\circ}$C in seven days. 
At this final temperature, the grown crystals are separated from the residual Bi-rich flux using a centrifugue (alumina wool is used as a filter) in a very rapid process, which also acts as a quench \cite{Zhao2018}. Well faceted ($\left\{ 100 \right\}$ and $\left\{ 111 \right\}$) mm-sized crystals of up to 1 gram are obtained, as seen in Fig. \ref{fig1}(a).  
If a slower cooling ramp is used (several weeks), smaller crystals are obtained. 

A typical XRD scan corresponding to the (hhh) planes is shown in Fig. \ref{fig1}(a). Additional peaks coming from Cu $K_{\beta}$ and W $L_{\alpha 1}$ and $L_{\alpha 2}$ (typical for copper tubes) are also observed for the very intense (222) and (333) reflections. Low intensity wide peaks around 40$^{\circ}$ and 60$^{\circ}$ come from the grease used to attach the crystal to the sample holder. An average (from several batches) lattice parameter $a =$ 6.696(5) \AA ~ is calculated, which agrees with the reported value \cite{Gao2017}.

Fig. \ref{fig1}(b) displays the magnetic susceptibility $\chi = M/H$ under an applied field $H$ = 5000 Oe. An almost temperature-independent diamagnetic response is observed, which could arise not only from the ion cores, but also from the conduction electrons (Landau diamagnetism) as in doped semiconductors, given the low electron effective masses reported in this compound. Comparable $\chi$ values are measured in the trigonal $\gamma$-PtBi$_2$ polymorph \cite{Xing2020}.

The temperature dependence of the electrical resistivity is depicted in Fig. \ref{fig1}(c). A metallic behavior is observed down to 2 K. The residual resistivity $\rho \left( 0 \right) \approx$ 0.065 $\mu \Omega\cdot$cm and the residual resistivity ratio $RRR = \rho(300 K)/\rho(0) \approx$ 515 are comparable to those reported in the literature \cite{Gao2017,Zhao2018}. Our samples quality is further confirmed by the magnetoresistance $MR = \left( R(H)-R(0) \right) / R(0)$ shown in the inset of Fig. \ref{fig1}(c), where at $T =$ 2 K, clear Shubnikov de Haas oscillations are observed at $\mu_0 H >$ 10 T. On the other hand, the $MR$ has a parabolic field dependence in the whole field range ($\mu_0 H =$ 16 T) and reaches a magnitude of $MR$ (9T) = 7700, about 46 \% of the highest reported value \cite{Gao2017}. 

Specific heat $c_p$ divided by temperature as a function of square temperature is shown in Fig. \ref{fig1}(d). The experimental data is fitted with the polynomial expression $c_p / T =\gamma + \beta T^2 + \epsilon T^4$ to give a Sommerfeld coefficient $\gamma$ = 2 mJ mol$^{-1}$ K$^{-2}$ and $\beta =$ 7.4 $\cdot$ 10$^{-4}$ J mol$^{-1}$ K$^{-4}$.
According to the Debye model $\beta$ = 234 3$N_a \, k_B \, \theta_D^{-3}$ ($N_a$ and $k_B$ are the Avogadro number and the Boltzmann constant, respectively), from which a Debye temperature $\theta_D =$ 199(1) K is obtained.
This value is considerable higher than in $\gamma$-PtBi$_2$ ($\theta_D =$ 145 K) \cite{Xu2016}. The layered structure of this trigonal structure, as compared with the highly compact cubic structure of $\beta$-PtBi$_2$ may explain this difference.

\subsection{$\beta$-Bi$_2$Pd}
\label{Pd}

The initial mixture (Bi:Pd = 2:1) is homogeneized at 900 $^{\circ}$C for 4 hours, then cooled down to 490 $^{\circ}$C in four days. From this temperature the still liquid mixture is further cooled down to 395 $^{\circ}$C in eight days. At this temperature the quartz capsule is quenched in cold water \cite{Herrera2015}. 
Large plateled-like crystals of several mm are obtained as seen in Fig. \ref{fig2}(a). The large face is parallel to (00l). A typical XRD scan corresponding to this family of planes is shown in Fig. \ref{fig2}(a). No spurious peaks other than Cu $K_{\beta}$ reflections are observed. Lattice parameters $c =$ 12.98(2) \AA ~ and $a =$ 3.369(6) \AA ~ are calculated from (00l) and (h0h) scans, in good agreement with previously reported values \cite{Imai2012,Herrera2015}. 

Fig. \ref{fig2}(b) displays the magnetic susceptibility $\chi$ under an applied field $H$ = 10000 Oe along the $a$- and $c$-axis. An almost temperature independent diamagnetic response is observed at all temperatures, in agreement with previous measurements \cite{Kolapo2019}. An important magnetic anisotropy can be seen as well.
The temperature dependence of the electrical resistivity is depicted in the main panel of Fig. \ref{fig2}(c). A metallic behavior is observed. The residual resistivity $\rho \left( 0 \right) \approx$ 23 $\mu \Omega\cdot$cm and $RRR \approx$ 3 are average among the reported values \cite{Kacmarcik2016,Matsuzaki2017}. At $T_c =$ 5.1 K, the system becomes superconducting (see lower inset of Fig. \ref{fig2}(c)). This is confirmed by zero-field-cooling magnetic moment measurements ($H =$ 3 Oe) that show a clear Meissner state below $T_c$ as shown in the upper inset of Fig. \ref{fig2}(c). Again, this critical temperature is average but lower than the higher reported value (= 5.4 K) \cite{Imai2012}.

Bulk superconductivity is supported by the specific heat as seen in Fig. \ref{fig2}(d). Zero-field $c_p$ shows a transition around 5 K that is suppressed by a moderate magnetic field ($H =$ 6 kOe). $c_p$(6 kOe) is fitted with the polynomial expression $c_p / T =\gamma + \beta T^2 + \epsilon T^4 + \delta T^6$ to give a Sommerfeld coefficient $\gamma$ = 7.37 mJ mol$^{-1}$ K$^{-2}$ and $\beta =$ 22.2 $\cdot$ 10$^{-4}$ J mol$^{-1}$ K$^{-4}$, from which a $\theta_D =$ 138(1) K is calculated. Both $\gamma$ and $\theta_D$ agree with reported values \cite{Imai2012,Matsuzaki2017,Pristas2018}. 
The electronic contribution to the specific heat, $c_p^{el}$, is obtained after subtracting the non-linear terms of the polynomial fit. $c_p^{el}$ is displayed in the inset of Fig. \ref{fig2}(d) showing a nearly compensated entropy at $T_c$.

\section{Thermal expansion}
\label{III}

\begin{figure*}[t]
\centering
\includegraphics[width=\textwidth]{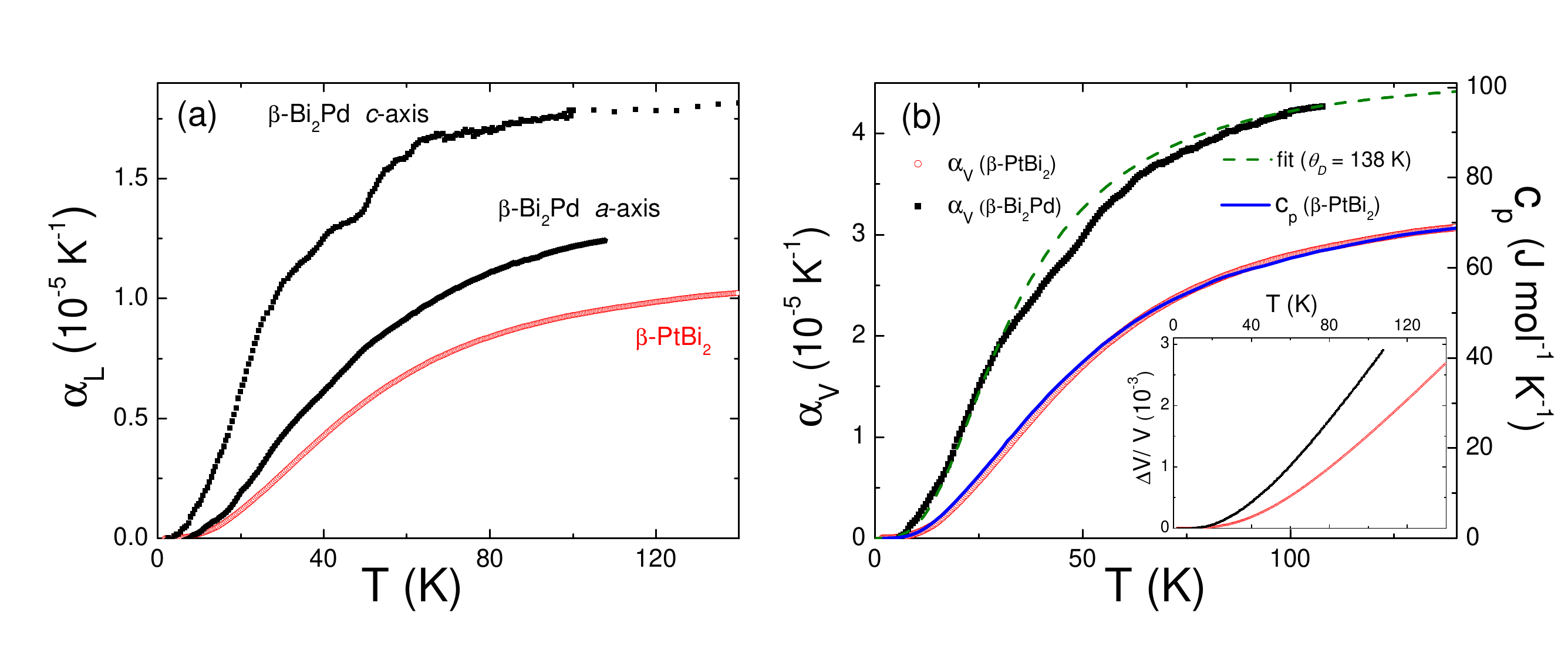}
\caption[]{(a) Temperature dependence of the linear thermal-expansion coefficients $\alpha_L$. (b) Temperature dependence of the volume thermal-expansion coefficients $\alpha_V$ (symbols). The solid curve corresponds to the experimental $c_p$ ($\beta$-PtBi$_2$) while the dash curve is a fit to $\alpha_V$($\beta$-Bi$_2$Pd) with the Debye model using a Debye temperature $\theta_D =$ 138 K . Inset: volume expansivity $\Delta V(T)  / V$.}
\label{fig3}
\end{figure*}

Thermal expansion was measured with a high-resolution ($\Delta L \leq$ 1 \AA) capacitive dilatometer \cite{Schmiedeshoff2006}. Several mm-sized samples were cut and polished along high-symmetry crystallographic directions.
Fig. \ref{fig3}(a) shows the temperature dependence of the linear-thermal-expansion coefficient $\alpha_L = \frac{1}{L}\left(\frac {\partial L}{\partial T} \right)_T$ for both $\beta$-PtBi$_2$ (measured along the [111] direction) and $\beta$-Bi$_2$Pd (measured along the two tetragonal axis, [100] and [001]). 
The dip around 50 K in $c$-axis $\alpha_L$($\beta$-Bi$_2$Pd) is associated with an experimental noisy condition rather than an intrinsic feature.  

The first thing to remark is that $\alpha_L$ is larger in $\beta$-Bi$_2$Pd along any direction than in $\beta$-PtBi$_2$. This gives rise to a sizeable difference in the volume thermal-expansion coefficient $\alpha_V = \frac{1}{V}\left(\frac {\partial V}{\partial T} \right)_T$ and the concomitant volume change of both compounds, as seen in Fig. \ref{fig3}(b) and its inset. Here, $\alpha_V$($\beta$-PtBi$_2$) = 3$\cdot \alpha_{111}$ and $\alpha_V$($\beta$-Bi$_2$Pd) = $\alpha_{001}$ + 2$\cdot \alpha_{100}$, given the cubic and tetragonal  crystal symmetry, respectively. 
The difference is interesting since both compounds have similar unit-cell volumes, $V_u$($\beta$-PtBi$_2$) = 75.06 \AA$^3$ and $V_u$($\beta$-Bi$_2$Pd) = 73.66 \AA$^3$.

$c_p$($\beta$-PtBi$_2$) is also shown in Fig. \ref{fig3}(b). The match with $\alpha_V$ is excellent in the whole temperature range indicating a nearly constant Gr\"uneisen parameter $\Gamma = \frac{N_a V_u B_s \alpha_V}{c_p}$, provided the isentropic bulk modulus $B_s$ has a negligible temperature dependence, as it is usual. 
Taking $B_s =$ 107-137.9 GPa from the literature \cite{Mangwejane2005,Chen2018}, $\Gamma$(100 K) = 2.19-2.83 is obtained in very close agreement with the expectation for lattice vibrations ($\approx$ 2) \cite{Moruzzi1988}.
We have not measured $c_p$($\beta$-Bi$_2$Pd) at high temperature. Instead, $\alpha_V$($\beta$-Bi$_2$Pd) is fitted with the expression for the specific heat given by the Debye model using $\theta_D =$ 138 K, obtained from the low-T $c_p$. The fit is good below 30 K and it overestimates the experimental $\alpha_V$ in the intermediate range (30 K $< T <$ 100 K).
Taking $c_p$ (100 K) = 68 J mol$^{-1}$ K$^{-1}$ and $B_s =$ 63.8 GPa from the literature \cite{Pristas2018,Zhou2019}, $\Gamma$(100 K) = 1.75 is calculated. 
The validity of the Debye fit is further confirmed by the fact that the ratio $\alpha_V / T$ has a parabolic temperature dependence below 6 K in both compounds (not shown here).  

No sign of the superconducting transition is observed in $\alpha_L$($\beta$-Bi$_2$Pd), even though the magnetization of the same sample shows a clear Meissner effect. From the Ehrenfest equation for second-order phase transitions, a jump in the thermal-expansion coefficient $\Delta \alpha_V =  \frac{dT}{dp}  \frac{\Delta c_p}{N_a V_u T_c}$ is expected. Here, $\frac{dT}{dp}$ = 2.5 $\cdot$ 10$^{-10}$ K Pa$^{-1}$ \cite{Pristas2018,Zhao2015} and $\Delta c_p$ = 5.7 $\cdot$ 10$^{-2}$ J mol$^{-1}$ K$^{-1}$ (from Fig. \ref{fig2}(d)) are the slope of the transition curve in the temperature-pressure ($T$ - $p$) phase diagram and the jump in the specific heat $c_p$ at the transition, respectively. From these experimental values a jump $\Delta \alpha_V =$ 6.3 $\cdot$ 10$^{-8}$  K$^{-1}$ is expected, which is comparable to the experimental resolution of our dilation experiments explaining why the transition is undetected. 

No magnetic field dependence of the lattice lengths was observed in any of both compounds up to $\mu_0 H =$ 16 T, which imposes an upper limit for the linear magnetostriction, $\left|\Delta L / L\right|$ (16 T) $<$ 4 $\cdot$ 10$^{-7}$. This fact confirms the insignificant electron-lattice coupling common to the non-magnetic TM.  

As mentioned above and due to this small electron-lattice coupling, expansivity data for TM candidates are scarce. In Table \ref{tab1} we compare our results with those from other Bi-based TI found in the literature. $\alpha_L$ ($\beta$-Bi$_2$Pd) shows a significant anisotropy between $\alpha_a$ and $\alpha_c$ (see Fig. \ref{fig3}(a)) as Bi$_2$Te$_3$ does, while $\alpha_V$($\beta$-Bi$_2$Pd) is larger than what is found in any other compound. 
On the other hand, $\alpha_V$($\beta$-PtBi$_2$) is considerable smaller and comparable to what is observed in Bi$_{2}$Se$_3$.  

\begin{table*}
\begin{tabular*}{\textwidth}{@{\extracolsep{\fill} } l c c c r}
														& Structure & Type & $\theta_D$ (K) & $\alpha$$\left[ \right.$100 K$\left. \right]$ (10$^{-5}$ K$^{-1}$)  \\
\hline
\\
\multirow{2}{*}{$\beta$-PtBi$_2$}  & \multirow{2}{*}{cubic} & \multirow{2}{*}{TSM} & \multirow{2}{*}{199$^a$} & $\alpha_a$ = 0.93$^a$  \\
																	 &							          &				               &	                        & $\alpha_V$ = 2.8$^a$  \\
\\
\multirow{3}{*}{$\beta$-Bi$_2$Pd}  & \multirow{3}{*}{tetragonal} & \multirow{3}{*}{TSC} & \multirow{3}{*}{138$^a$}  & $\alpha_a$ = 1.22$^a$  \\
																	 &							               &				              &                           & $\alpha_c$ = 1.78$^a$  \\
																	 &							               &				              &	                          & $\alpha_V$ = 4.2$^a$  \\
\\
\multirow{3}{*}{Bi$_{1-x}$Sb$_x$}  & \multirow{3}{*}{rhombohedral} & \multirow{3}{*}{TI} &                      & $\alpha_a \lesssim$ 1.2$^b$  \\
																	 &							                 &				             &                      & $\alpha_c \gtrsim$ 1.2$^b$  \\
																	 &							                 &				             &	                    & $\alpha_V \sim$ 3.5$^b$  \\
\\
\multirow{3}{*}{Bi$_{2}$Se$_3$}    & \multirow{3}{*}{rhombohedral} & \multirow{3}{*}{TI} & \multirow{3}{*}{182.3$^c$} & $\alpha_a$ = 0.7$^d$  \\
																	 &							                 &				              &                            & $\alpha_c$ = 1.2$^d$  \\
																	 &							                 &				              &	                           & $\alpha_V$ = 2.6$^d$  \\
\\
\multirow{3}{*}{Bi$_{2}$Te$_3$}    & \multirow{3}{*}{rhombohedral} & \multirow{3}{*}{TI} & \multirow{3}{*}{164.9$^c$} & $\alpha_a$ = 0.9$^e$  \\
																	 &							                 &				              &                            & $\alpha_c$ = 1.9$^e$  \\
																	 &							                 &				              &	                           & $\alpha_V$ = 3.7$^e$  \\
\end{tabular*}
	\caption{Debye temperature $\theta_D$ and thermal-expansion coefficients $\alpha$ for Bi-based topological material candidates. Superscript letters correspond to References: a = this work, b = \cite{Malik2016}, c = \cite{Madelung2004}, d = \cite{Chen2011}, e = \cite{Pavlova2011}}
	\label{tab1}
\end{table*}

\section{Conclusions}
\label{IV}

Low-temperature thermal-expansion ($T <$ 120 K) has been measured in single crystals of cubic $\beta$-PtBi$_2$ and tetragonal $\beta$-Bi$_2$Pd. Both systems have received much attention in recent years due to the potential topological nature of their electronic properties.

The linear thermal-expansion coefficient of superconducting $\beta$-Bi$_2$Pd shows a pronounced anisotropy between the $a$- and $c$-axis. The volume thermal-expansion coefficient $\alpha_V$ of $\beta$-Bi$_2$Pd is considerable larger than that of the semimetal $\beta$-PtBi$_2$. 
The coefficient $\alpha_V$($\beta$-PtBi$_2$) nearly matches the experimental specific heat, from which a Debye temperature $\theta_D =$ 199 K is obtained. On the other hand, $\alpha_V$($\beta$-Bi$_2$Pd) reasonably fits the Debye model with $\theta_D =$ 138 K, extracted from the low-temperature specific heat.

An almost temperature independent Gr\"uneisen parameter $\Gamma \approx$ 2 is obtained for both compounds in good agreement with the expectation for lattice vibrations.
No magnetostriction is observed in any of both compounds up to $\mu_0 H =$ 16 T.

\section{Acknowledgments}

This work was supported by CONICET grant number PIP2021-11220200101796CO, ANPCyT grant number PICT2019-02396, Universidad Nacional de Cuyo (SIIP) grant number 06/C559. 
P.P. acknowledges support from Deutsche Forschungsgemeinschaft (DFG) through SFB 1143 (project id 247310070).
N.H. is member of the INN, CNEA-CONICET, Argentina.




\begin{thebibliography}{99}

\bibitem{Hasan2010} M. Z. Hasan, C. L. Kane, Colloquium: Topological insulators, Rev. Mod. Phys. 82 (2010) 3045-3067.

\bibitem{Qi2011} X.-L. Qi, S.-C. Zhang, Topological insulators and superconductors, Rev. Mod. Phys. 83 (2011) 1057-1110.

\bibitem{Sato2017} M. Sato, Y. Ando, Topological superconductors: a review, Rep. Prog. Phys. 80 (2017) 076501.

\bibitem{Yan2017} B. Yan, C. Felser, Topological Materials: Weyl Semimetals, Annu. Rev. Condens. Matter Phys. 8 (2017) 337–354.

\bibitem{Armitage2018} N. P. Armitage, E. J. Mele, A. Vishwanath, Weyl and Dirac semimetals in three-dimensional solids, Rev. Mod. Phys. 90 (2018) 015001.

\bibitem{Leijnse2012} M. Leijnse, K. Flensberg, Introduction to topological superconductivity and Majorana fermions, Semicond. Sci. Technol. 27 (2012) 124003.

\bibitem{Bernevig2006} B. A. Bernevig, T. L. Hughes, S. C. Zhang, Quantum spin Hall effect and topological phase transition in HgTe quantum wells, Science 314 (2006) 1757-1761.

\bibitem{Konig2007} M. K\"onig, S. Wiedmann, C. Br\"une, A. Roth, H. Buhmann, L. Molenkamp, X.-L. Qi, S.-C. Zhang, Quantum spin Hall insulator state in HgTe quantum wells, Science 318 (2007) 766-770.

\bibitem{Hsieh2008} D. Hsieh, D. Qian, L. Wray, Y. Xia, Y. S. Hor, R. J. Cava, M. Z. Hasan, A topological Dirac insulator in a quantum spin Hall phase, Nature 452 (2008) 970-974.

\bibitem{Zhang2009} H. Zhang, C. X. Liu, X. L. Qi, X. Dai, Z. Fang, S. C. Zhang, Topological insulators in Bi$_2$Se$_3$, Bi$_2$Te$_3$ and Sb$_2$Te$_3$ with a single Dirac cone on the surface, Nat. Phys. 5 (2009) 438-442.

\bibitem{Xia2009} Y. Xia, D. Qian, D. Hsieh, L. Wray, A. Pal, H. Lin, A. Bansil, D. Grauer, Y. S. Hor, R. J. Cava, M. Z. Hasan, Observation of a large-gap topological-insulator class with a single Dirac cone on the surface, Nat. Phys. 5 (2009) 398-402.

\bibitem{Chadov2010} S. Chadov, X. L. Qi, J. K\"ubler, G.H. Fecher, C. Felser, S. C. Zhang, Tunable multifunctional topological insulators in ternary Heusler compounds, Nat. Mater. 9 (2010) 541-545.

\bibitem{Lin2010} H. Lin, L. A. Wray, Y. Xia, S. Xu, S. Jia, R. J. Cava, A. Bansil, M. Z. Hasan, Half-Heusler ternary compounds as new multifunctional experimental platforms for topological quantum phenomena, Nature Mater. 9 (2010) 546-549.

\bibitem{Xiao2010} D. Xiao, Y. Yao, W. Feng, J. Wen, W. Zhu, X. Chen, G. M. Stocks, Z. Zhang, Half-Heusler compounds as a new class of three-dimensional topological insulators, Phys. Rev. Lett. 105 (2010) 096404.

\bibitem{Ali2014} M. N. Ali, J. Xiong, S. Flynn, J. Tao, Q. D. Gibson, L. M. Schoop, T. Liang, N. Haldolaarachchige, M. Hirschberger, N. P. Ong, R. J. Cava, Large, non-saturating magnetoresistance in WTe$_2$, Nature 514 (2014) 205–208.

\bibitem{Sun2015} Y. Sun, S. Wu, M. N. Ali, C. Felser, B. Yan, Prediction of Weyl semimetal in orthorhombic MoTe$_2$, Phys. Rev. B 92 (2015) 161107(R).

\bibitem{Gao2017} W. Gao, N. Hao, F. Zheng, W. Ning, M. Wu, X. Zhu, G. Zheng, J. Zhang, J. Lu, H. Zhang, C. Xi, J. Yang, H. Du, P. Zhang, Y. Zhang, M. Tian, Extremely large magnetoresistance in a topological semimetal candidate pyrite PtBi$_2$, Phys. Rev. Lett. 118 (2017) 256601.

\bibitem{Lv2015} B. Q. Lv, H. M. Weng, B. B. Fu, X. P. Wang, H. Miao, J. Ma, P. Richard, X. C. Huang, L. X. Zhao, G. F. Chen, Z. Fang, X. Dai, T. Qian, H. Ding, Experimental discovery of Weyl semimetal TaAs, Phys. Rev. X 5 (2015) 031013.

\bibitem{Xu2015} S. Xu, I. Belopolski, D. S. Sanchez, C. Zhang, G. Chang, C. Guo, G. Bian, Z. Yuan, H. Lu, T. Chang, P. P. Shibayev, M. L. Prokopovych, N. Alidoust, H. Zheng, C. Lee, S. Huang, R. Sankar, F. Chou, C. Hsu, H. Jeng, A. Bansil, T. Neupert, V. N. Strocov, H. Lin, S. Jia, M. Z. Hasan, Experimental discovery of a topological Weyl semimetal state in TaP, Sci. Adv. 1 (2015) 1501092.

\bibitem{Fang2019} H. F. Yang, L. X. Yang, Z. K. Liu, Y. Sun, C. Chen, H. Peng, M. Schmidt, D. Prabhakaran, B. A. Bernevig, C. Felser, B. H. Yan, Y. L. Chen, Topological Lifshitz transitions and Fermi arc manipulation in Weyl semimetal NbAs, Nat. Commun. 10 (2019) 3478.

\bibitem{Shekhar2015} C. Shekhar, A. K. Nayak, Y. Sun, M. Schmidt, M. Nicklas, I. Leermakers, U. Zeitler, Y. Skourski, J. Wosnitza, Z. Liu, Y. Chen, W. Schnelle, H. Borrmann, Y. Grin, C. Felser, B. Yan, Extremely large magnetoresistance and ultrahigh mobility in the topological Weyl semimetal candidate NbP, Nature Phys. 11 (2015) 645–649.

\bibitem{Liu2014} Z. K. Liu, J. Jiang, B. Zhou, Z. J. Wang, Y. Zhang, H. M. Weng, D. Prabhakaran, S-K. Mo, H. Peng, P. Dudin, T. Kim, M. Hoesch, Z. Fang, X. Dai, Z. X. Shen, D. L. Feng, Z. Hussain, Y. L. Chen, A stable three-dimensional topological Dirac semimetal Cd$_3$As$_2$, Nature Mater. 13 (2014) 677–681.

\bibitem{Mackenzie2013} A. P. Mackenzie, Y. Maeno, The superconductivity of Sr$_2$RuO$_4$ and the physics of spin-triplet pairing, Rev. Mod. Phys. 75 (2003) 657.

\bibitem{Hor2010} Y. S. Hor, A. J. Williams, J. G. Checkelsky, P. Roushan, J. Seo, Q. Xu, H. W. Zandbergen, A. Yazdani, N. P. Ong, R. J. Cava, Superconductivity in Cu$_x$Bi$_2$Se$_3$ and its implications for pairing in the undoped topological insulator, Phys. Rev. Lett. 104 (2010) 057001.

\bibitem{Sasaki2012} S. Sasaki, Z. Ren, A. A. Taskin, K. Segawa, L. Fu, Y. Ando, Odd-parity pairing and topological superconductivity in a strongly spin-orbit coupled semiconductor,
Phys. Rev. Lett. 109 (2012) 217004.
 
\bibitem{Benia2016} H. M. Benia, E. Rampi, C. Trainer, C. M. Yim, A. Maldonado, D. C. Peets, A. Stöhr, U. Starke, K. Kern, A. Yaresko, G. Levy, A. Damascelli, C. R. Ast, A. P. Schnyder, P. Wahl, Observation of Dirac surface states in the noncentrosymmetric superconductor BiPd, Phys. Rev. B 94 (2016) 121407(R).

\bibitem{Shang2020} T. Shang, J. Z. Zhao, D. J. Gawryluk, M. Shi, M. Medarde, E. Pomjakushina, T. Shiroka, Superconductivity and topological aspects of the rocksalt carbides NbC and TaC, Phys. Rev. B 101 (2020) 214518.

\bibitem{Wang2016} K.L. Wang, M. Lang, X. Kou, Spintronics of topological insulators. In:  Y. Xu, D. Awschalom, J. Nitta (eds) Handbook of Spintronics (Springer, Dordrecht, 2016).

\bibitem{Imai2012} Y. Imai, F. Nabeshima, T. Yoshinaka, K. Miyatani, R. Kondo, S. Komiya, I. Tsukada, A. Maeda, Superconductivity at 5.4 K in $\beta$-Bi$_2$Pd, J. Phys. Soc. Jpn. 81 (2012) 113708.

\bibitem{Haberkorn2022} N. Haberkorn, V. F. Correa, unpublished.

\bibitem{Zhao2018} L. Zhao, L. Xu, H. Zuo, X. Wu, G. Gao, Z. Zhu, Fermi surface and carrier compensation of pyrite-type PtBi$_2$ revealed by quantum oscillations, Phys. Rev. B 98 (2018) 085137.

\bibitem{Xing2020} L. Xing, R. Chapai, R. Nepal, R. Jin, Topological behavior and Zeeman splitting in trigonal PtBi$_{2-x}$ single crystals, npj Quantum Mater. 5 (2020) 10.

\bibitem{Xu2016} C. Q. Xu, X. Z. Xing, X. Xu, B. Li, B. Chen, L. Q. Che, X. Lu, J. Dai, Z. X. Shi, Synthesis, physical properties, and band structure of the layered bismuthide PtBi$_2$, Phys. Rev. B 94 (2016) 165119. 

\bibitem{Herrera2015} E. Herrera, I. Guillam\'on, J. A. Galvis, A. Correa, A. Fente, R. F. Luccas, F. J. Mompean, M. Garc\'ia-Hern\'andez, S. Vieira, J. P. Brison, H. Suderow, Magnetic field dependence of the density of states in the multiband superconductor $\beta$-Bi$_2$Pd, Phys. Rev. B 92 (2015) 054507.

\bibitem{Kolapo2019} A. Kolapo, T. Li, P. Hosur, J. H. Miller Jr., Possible transport evidence for three-dimensional topological
superconductivity in doped $\beta$-PdBi$_2$, Sci. Rep. 9 (2019) 12504.

\bibitem{Kacmarcik2016} J. Ka\v cmar\v c\'ik, Z. Pribulov\'a, T. Samuely, P. Szab\'o, V. Cambel, J. \v Solt\'ys, E. Herrera, H. Suderow, A. Correa-Orellana, D. Prabhakaran, P. Samuely, Single-gap superconductivity in $\beta$-Bi$_2$Pd, Phys. Rev. B 93 (2016)144502. 

\bibitem{Matsuzaki2017} H. Matsuzaki, K. Nagai, N. Kase, T. Nakano, N. Takeda, Superconducting gap of the single crystal $\beta$-Bi$_2$Pd, Journal of Physics: Conf. Series 871 (2017) 012004.

\bibitem{Pristas2018} G. Prist\'a\v s, M. Orend\'a\v c, S. Gab\'ani, J. Ka\v cmar\v c\'ik, E. Ga\v zo, Z. Pribulov\'a, A. Correa-Orellana, E. Herrera, H. Suderow, P. Samuely, Pressure effect on the superconducting and the normal state of $\beta$-Bi$_2$Pd, Phys. Rev. B 97 (2018) 134505.

\bibitem{Schmiedeshoff2006} G. M. Schmiedeshoff, A. W. Lounsbury, D. J. Luna, S. J. Tracy, A. J. Schramm, S. W. Tozer, V. F. Correa, S. T. Hannahs, T. P. Murphy, E. C. Palm, A. H. Lacerda, S. L. Bud'ko, P. C. Canfield, J. L. Smith, J. C. Lashley, J. C. Cooley, Versatile and compact capacitive dilatometer, Rev. Sci. Instrum. 77 (2006) 123907.

\bibitem{Mangwejane2005} S. S. Mangwejane, Master Thesis (University of Limpopo, South Africa, 2005).

\bibitem{Chen2018} X. Chen, D. Shao, C. Gu, Y. Zhou, C. An, Y. Zhou, X. Zhu, T. Chen, M. Tian, J. Sun, Z. Yang, Pressure-induced multiband superconductivity in pyrite PtBi$_2$ with perfect electron-hole compensation, Phys. Rev. Materials 2 (2018) 054203.

\bibitem{Moruzzi1988} V. L. Moruzzi, J. F. Janak, K. Schwarz, Calculated thermal properties of metals, Phys. Rev. B 37 (1988) 790-799.

\bibitem{Zhou2019} Y. Zhou, X. Chen, C. An, Y. Zhou, L. Ling, J. Yang, C. Chen, L. Zhang, M. Tian, Z. Zhang, Z. Yang, Pressure-induced irreversible evolution of superconductivity in Bi$_2$Pd, Phys. Rev. B 99 (2019) 054501. 

\bibitem{Zhao2015} K. Zhao, B. Lv, Y. Xue, X. Zhu, L. Z. Deng, Z. Wu, C. W. Chu, Chemical doping and high-pressure studies of layered $\beta$-PdBi$_2$ single crystals, Phys. Rev. B 92 (2015) 174404.

\bibitem{Malik2016} K. Malik, D. Das, A. K. Deb, V. A. Kulbachinskii, V. Srihari, S. Bandyopadhyay, A. Banerjee, Evidence of iso-structural phase transition in rhombohedral Bi-Sb alloy, EPL 115 (2016) 58001.

\bibitem{Madelung2004} O. Madelung, Semiconductors: Data Handbook, 3rd ed. (Springer, Berlin, 2004).

\bibitem{Chen2011} X. Chen, H. D. Zhou, A. Kiswandhi, I. Miotkowski, Y. P. Chen, P. A. Sharma, A. L. Lima Sharma, M. A. Hekmaty, D. Smirnov, Z. Jiang, Thermal expansion coefficients of Bi$_{2}$Se$_3$ and Sb$_2$Te$_3$ crystals from 10 K to 270 K, Appl. Phys. Lett. 99 (2011) 261912.

\bibitem{Pavlova2011} L. M. Pavlova, Yu. I. Shtern, R. E. Mironov, Thermal expansion of bismuth telluride, High Temp. 49 (2011) 369–379.









\end{thebibliography}


\end{document}